\documentclass[12pt]{article}
\usepackage{amsmath}
\usepackage{graphicx}
\usepackage{bm}
\usepackage{fancyhdr}
\usepackage{amssymb}
\usepackage{setspace}
\usepackage{color}
\renewcommand{\theequation}{\arabic{section}.\arabic{equation}}

\begin{document}



\def\a{\alpha}
\def\b{\beta}
\def\d{\delta}
\def\e{\epsilon}
\def\g{\gamma}
\def\h{\mathfrak{h}}
\def\k{\kappa}
\def\l{\lambda}
\def\o{\omega}
\def\p{\wp}
\def\r{\rho}
\def\t{\tau}
\def\s{\sigma}
\def\z{\zeta}
\def\x{\xi}
\def\V={{{\bf\rm{V}}}}
 \def\A{{\cal{A}}}
 \def\B{{\cal{B}}}
 \def\C{{\cal{C}}}
 \def\D{{\cal{D}}}
\def\K{{\cal{K}}}
\def\O{\Omega}
\def\R{\bar{R}}
\def\T{{\cal{T}}}
\def\L{\Lambda}
\def\f{E_{\tau,\eta}(sl_2)}
\def\E{E_{\tau,\eta}(sl_n)}
\def\Zb{\mathbb{Z}}
\def\Cb{\mathbb{C}}

\def\R{\overline{R}}

\def\beq{\begin{equation}}
\def\eeq{\end{equation}}
\def\bea{\begin{eqnarray}}
\def\eea{\end{eqnarray}}
\def\ba{\begin{array}}
\def\ea{\end{array}}
\def\no{\nonumber}
\def\le{\langle}
\def\re{\rangle}
\def\lt{\left}
\def\rt{\right}

\newtheorem{Theorem}{Theorem}
\newtheorem{Definition}{Definition}
\newtheorem{Proposition}{Proposition}
\newtheorem{Lemma}{Lemma}
\newtheorem{Corollary}{Corollary}
\newcommand{\proof}[1]{{\bf Proof. }
        #1\begin{flushright}$\Box$\end{flushright}}

\baselineskip=20pt

\newfont{\elevenmib}{cmmib10 scaled\magstep1}
\newcommand{\preprint}{
   \begin{flushleft}
   \end{flushleft}\vspace{-1.3cm}
   \begin{flushright}\normalsize
   \end{flushright}}
\newcommand{\Title}[1]{{\baselineskip=26pt
   \begin{center} \Large \bf #1 \\ \ \\ \end{center}}}
\newcommand{\Author}{\begin{center}
   \large \bf
Yi Qiao${}^{a,b}$,~Xin Zhang${}^{c}$,~Kun Hao${}^{a,b}$,~Junpeng Cao${}^{c,d,e}$,~Guang-Liang Li${}^{f}$,~Wen-Li Yang${}^{a,b}\footnote{Corresponding author:
wlyang@nwu.edu.cn}$ and~ Kangjie Shi${}^{a,b}$

 \end{center}}
\newcommand{\Address}{\begin{center}

     ${}^a$Institute of Modern Physics, Northwest University,
     Xian 710069, China\\
     ${}^b$Shaanxi Key Laboratory for Theoretical Physics Frontiers,  Xian 710069, China\\
     ${}^c$Beijing National Laboratory for Condensed Matter
           Physics, Institute of Physics, Chinese Academy of Sciences, Beijing
           100190, China\\
     ${}^d$Collaborative Innovation Center of Quantum Matter, Beijing,
     China\\
     ${}^e$School of Physical Sciences, University of Chinese Academy of Sciences, Beijing, China\\
     ${}^f$Department of Applied Physics, Xian Jiaotong University, Xian 710049, China\\

   \end{center}}

\preprint \thispagestyle{empty}
\bigskip\bigskip\bigskip

\Title{A convenient basis for the Izergin-Korepin model} \Author

\Address \vspace{0.1cm}

\begin{abstract}
We propose a convenient  orthogonal basis of the Hilbert space for the Izergin-Korepin model (or the quantum spin chain
associated with the $A^{(2)}_{2}$ algebra). It is shown that the monodromy-matrix elements acting
on the basis take relatively simple forms (c.f. acting on the original basis ), which is quite similar as that in
the so-called F-basis for the quantum spin chains associated with $A$-type (super)algebras. As an application, we present
the  recursive expressions of Bethe states in the basis for the Izergin-Korepin model.
\vspace{1truecm}

\end{abstract}

\newpage


\hbadness=10000

\tolerance=10000

\hfuzz=150pt

\vfuzz=150pt

\section{Introduction}
\label{intro} \setcounter{equation}{0}

The quantum inverse scattering method (QISM) (or the algebraic Bethe Ansatz method (ABA)) provides a
powerful method of solving eigenvalue problems for quantum integrable systems \cite{Kor93}. In this framework,
the quasi-particle creation and annihilation operators are constructed by the off-diagonal matrix elements of
the monodromy-matrix. The Bethe states (eigenstates) are obtained by acting the creation operators on
a reference state \cite{Kor93, Wan15}. However, the apparently simple action of creation operators is intricate on the level
of the local operators by non-local effects arising from polarization clouds or compensating
exchange terms \cite{Mai96}. This makes the exact and explicit computation of correlation functions
very involved (if not impossible). It was shown \cite{Mai96} that for the inhomogeneous XXX and XXZ spin chains there
does exist a particular basis (the so-called F-basis \cite{Dri83}), in which the actions of the  monodromy
matrices can be simplified dramatically. This leads to  the analysis of these models  in the F-basis \cite{Kit99}.  Since then such
a basis has been constructed  for other models only related to the $A$-type algebras: the high-spin XXX spin chains  \cite{Ter99}, the quantum integrable spin
chains \cite{Alb00} associated with $gl(m)$ algebra and their elliptic generalizations \cite{Alb00-1,Alb01}, and the supersymmetric Fermionic models related
to the superalgebras $gl(m|n)$ \cite{Yan04,Zha06}. Whether this kind of basis does exist for other quantum integrable systems (especially for those related to the non $A$-type
(super)algebras) is still an interesting open problem. The aim of this paper focuses on this problem for the first simplest quantum spin chain beyond $A$-type,
namely, the Izergin-Korepin (IK) model \cite{Ize81}.

The IK model has played a fundamental role in
quantum integrable models associated with algebras beyond $A$-type. It was introduced as a quantum integrable model related
to the Dodd-Bullough-Mikhailov or Jiber-Mikhailov-Shabat model \cite{Dod77,Zhi79}, one of two
integrable relativistic models containing one scalar field (the other is sine-Gordon model).
The $R$-matrix of the model corresponds to the simplest twisted affine algebra $A^{(2)}_2$. Moreover,
it also has many applications in the studies of  the loop models \cite{Yun95-1} and
self-avoiding walks  \cite{Bat95}.  The Bethe Ansatz solution for  eigenvalues of the IK model with the periodic boundary condition
was first given by  Reshetikhin  with his elegant analytical Bethe Ansatz method \cite{Res83}. The corresponding Bethe states
was then constructed by  Tarasov \cite{Tar88}, which initiated the way to construct  Bethe states for quantum integrable models
beyond $A$-type \cite{Yun95-1,Tar88,Mat95,Fan97,Lim99,Hou99, Li03}. The purpose of the present paper is to propose a representation basis for the IK model with periodic boundary condition, which would play a similar role as that of the F-basis for quantum integrable systems related to the $A$-type.

The paper is organized as follows. Section 2 serves as an introduction to our notations for
the IK model with the periodic boundary condition. In section 3, we propose an orthogonal basis of the Hilbert space of the model.
It is shown that the matrix elements of the monodromy matrix acting
on this basis take simple forms, comparing with those in the original basis. In section 4, we give the  recursive relations of the vector components of Bethe states in
this basis,  which can determine the explicit expressions of the states. We give the solution of the quantum inverse scattering problem
for the IK model.\footnote{The general method to solve the quantum inverse problem for an integrable spin chain was given in \cite{Goh00,Mai00}. Here we
just list the results for this particular  model.}
The concluding remarks are given in section 5. Some detailed technical calculations are given in Appendices A-C.

\section{IK model}
\setcounter{equation}{0}
Throughout, ${\rm\bf V}$ denotes a three-dimensional linear space with an orthonormal basis $\{|i\rangle|i=1,2,3\}$.
We shall adopt the standard notations: for any matrix $A\in {\rm End}({\rm\bf V})$, $A_j$ is an
embedding operator in the tensor space ${\rm\bf V}\otimes
{\rm\bf V}\otimes\cdots$, which acts as $A$ on the $j$-th space and as
identity on the other factor spaces; For $B\in {\rm End}({\rm\bf V}\otimes {\rm\bf V})$, $B_{ij}$ is an embedding
operator of $B$ in the tensor space, which acts as identity
on the factor spaces except for the $i$-th and $j$-th ones.

The $R$-matrix $R(u)\in {\rm End}({\rm\bf V}\otimes {\rm\bf V})$ of
the IK model is given by \cite{Ize81}
 \bea
 R_{12}(u)=\left(\begin{array}{r|r|r}{\begin{array}{rrr}c(u)&&\\&b(u)&\\&&d(u)\end{array}}
           &{\begin{array}{lll}&&\\e(u)&{\,\,\,\,\,\,\,\,\,\,}&{\,\,\,\,\,\,\,\,\,\,}\\{\,\,\,\,\,\,}&g(u)&{\,\,\,\,\,\,}\end{array}}
           &{\begin{array}{lll}&&\\&&\\f(u)&{\,\,\,\,\,\,\,\,\,\,}&{\,\,\,\,\,\,\,\,\,\,}\end{array}}\\[12pt]
 \hline {\begin{array}{rrr}&\bar{e}(u)&\\&&\bar{g}(u)\\&&\end{array}}&
           {\begin{array}{ccc}b(u)&&\\&a(u)&\\&&b(u)\end{array}}
           &{\begin{array}{lll}&&\\g(u)&{\,\,\,\,\,\,\,\,\,\,}&{\,\,\,\,\,\,\,\,\,\,}\\&e(u)&{\,\,\,\,\,\,\,\,\,\,}\end{array}}\\[12pt]
 \hline {\begin{array}{ccc}&&\bar{f}(u)\\&&\\&&\end{array}}
           &{\begin{array}{ccc}&\bar{g}(u)&\\&&\bar{e}(u)\\&&\end{array}}
           &{\begin{array}{ccc}d(u)&&\\&b(u)&\\&&c(u)\end{array}} \end{array}\right),\label{R-matrix}
\eea
\noindent where the matrix elements are \bea
&&a(u)=\sinh(u\hspace{-0.04truecm}-\hspace{-0.04truecm}3\eta)
\hspace{-0.04truecm}-\hspace{-0.04truecm}\sinh
5\eta\hspace{-0.04truecm}+\hspace{-0.04truecm} \sinh
3\eta\hspace{-0.04truecm}+\hspace{-0.04truecm}\sinh\eta,\,\,
b(u)=\sinh(u\hspace{-0.04truecm}-\hspace{-0.04truecm}3\eta)
\hspace{-0.04truecm}+\hspace{-0.04truecm}\sinh3\eta,\no\\[6pt]
&&c(u)=\sinh(u-5\eta)+\sinh\eta,\quad d(u)=\sinh(u-\eta)+\sinh\eta,\no\\[6pt]
&&e(u)=-2e^{-\frac{u}{2}}\sinh2\eta\cosh(\frac{u}{2}-3\eta),\quad \bar{e}(u)=-2e^{\frac{u}{2}}\sinh2\eta\cosh(\frac{u}{2}-3\eta),\no\\[6pt]
&&f(u)=-2 e^{-u+2\eta}\sinh\eta\sinh2\eta-e^{-\eta}\sinh4\eta,\no\\[6pt]
&&\bar{f}(u)=2 e^{u-2\eta}\sinh\eta\sinh2\eta-e^{\eta}\sinh4\eta,\no\\[6pt]
&&g(u)=2e^{-\frac{u}{2}+2\eta}\sinh\frac{u}{2}\sinh 2\eta,\quad \bar{g}(u)=-2e^{\frac{u}{2}-2\eta}\sinh\frac{u}{2}\sinh 2\eta.
\label{R-element-2}
\eea
The $R$-matrix satisfies the quantum Yang-Baxter equation (QYBE)
\bea
R_{12}(u_1-u_2)R_{13}(u_1-u_3)R_{23}(u_2-u_3)
=R_{23}(u_2-u_3)R_{13}(u_1-u_3)R_{12}(u_1-u_2).\label{QYB}
\eea
For convenience, in the following parts of this paper, let us introduce some functions
\bea
&&\hspace{-1.2truecm}\omega(u)=\frac{c(u)d(u)}{a(u)d(u)-g(u)\bar g(u)},\quad
y(u)=\frac{d(u)}{\bar g(u)},\quad
\bar y(u)=\frac{d(u)}{g(u)},\quad
z(u)=\frac{c(u)}{b(u)}.\label{Functions}
\eea
The monodromy-matrix
$T(u)$ is an $n\times n$ matrix with operator-valued elements
acting on ${\rm\bf V}^{\otimes N}$ as
\bea
T_0(u)=R_{0N}(u-\theta_N)R_{0\,N-1}(u-\theta_{N-1})\cdots
R_{01}(u-\theta_1),\label{Monodromy-1}
\eea
where $\{\theta_j|j=1,\cdots,N\}$ are generic free complex parameters
which are usually called the inhomogeneous parameters. The QYBE (\ref{QYB}) implies that
the monodromy-matrix $T(u)$ satisfies the exchange relations (or the Yang-Baxter relations)
\bea
R_{12}(u-v)\,T_1(u)\,T_2(v)= T_2(v)\,T_1(u)\,R_{12}(u-v).\label{RLL}
\eea
The corresponding transfer matrix $t(u)$  can be constructed by the standard way \cite{Kor93} as
\begin{eqnarray}
t(u)=tr_0\lt\{T_0(u)\rt\}.\label{transfer}
\end{eqnarray}
The IK model  with periodic boundary condition is a quantum spin chain described by
the Hamiltonian
\bea
H=\frac{\partial}{\partial u}\lt.\lt\{\ln t(u)\rt\}\rt|_{u=0,\{\theta_i\}=0}=\frac{1}{\sinh\eta-\sinh5\eta}\sum_{i=1}^NH_{i,i+1},\label{Ham}
\eea
where the local Hamiltonian  $H_{i,i+1}$ is
\bea
H_{i,i+1}= \frac{\partial}{\partial u}\lt.\lt\{P_{i,i+1} R_{i,i+1}(u)\rt\}\rt|_{u=0}.
\eea
The periodic boundary condition for the Hamiltonian (\ref{Ham}) reads
\bea
H_{N,N+1}=H_{N,1}.\label{Periodic-boundary}
\eea
The QYBE leads to the fact that the transfer matrices $t(u)$ given by (\ref{transfer}) with
different spectral parameters are mutually commuting:
\begin{equation}\label{t-commu}
  [t(u),\,t(v)]=0.\no
\end{equation}
This ensures the integrability of
the IK model with periodic boundary described by the Hamiltonian (\ref{Ham}) and (\ref{Periodic-boundary}).

\section{Orthogonal basis for the IK model}
\setcounter{equation}{0}
It was shown \cite{Mai96} that for the inhomogeneous XXX and XXZ spin chains there
does exist a particular basis (the so-called F-basis \cite{Dri83}), in which the actions of the  monodromy
matrices can be simplified dramatically. Since then such a basis has been constructed  for other models only
related to the $A$-type algebras \cite{Ter99, Alb00, Alb00-1, Alb01, Yan04, Zha06}.
This leads to  the the F-basis analysis of these models \cite{Kit99, Zha06}.

In this section, we propose a convenient basis of the Hilbert space parameterized by the $N$ generic inhomogeneity parameters
$\{\theta_j|j=1,\cdots,N\}$. It is found that the actions of monodromy-matrix elements on this basis
take drastically simple forms like those in the so-called F-basis \cite{Mai96,Dri83,Alb00,Alb01} for the models related to the A-type (super)algebras.
For convenience, let us introduce the notations
\bea
&&A_i(u)=T^i_i(u),\quad B_1(u)=T^1_2(u),\quad B_2(u)=T^1_3(u),\quad B_3(u)=T^2_3(u), \quad {\rm for}\,\, i=1,2,3,\label{AB-operator}\no
\\&&C_1(u)=T^2_1(u),\quad C_2(u)=T^3_1(u), \quad C_3(u)=T^3_2(u).\label{C-operator}
\eea
The monodromy-matrix becomes
\bea
 T(u)=
\left(
  \begin{array}{ccc}
    A_1(u) & B_1(u) & B_2(u)\\
    C_1(u) & A_2(u) & B_3(u)\\
    C_2(u) & C_3(u) & A_3(u)\\
  \end{array}
\right).\label{T-matrix}
\eea
These operators satisfy the quadratic commutation relation (\ref{RLL}) (or the Yang-Baxter algebra) whose structure constants are given by the matrix elements of the $R$-matrix. The commutation relation  allows us to derive the exchange relations among the operators in (\ref{T-matrix}). Some relevant exchange relations for our purpose among the operators are given in Appendix A.

Let us introduce the left quasi-vacuum state $\langle 0|$ and the right quasi-vacuum state $ |0\rangle$ as follows
\bea
\langle 0|=\left(
  \begin{array}{ccc}
    1, & 0, & 0\\
  \end{array}
\right)_{[1]}\otimes\cdots\otimes\left(
  \begin{array}{ccc}
    1, & 0, & 0\\
  \end{array}
\right)_{[N]},\quad\quad|0\rangle=\left(\begin{array}{c}
    1 \\
    0 \\
    0 \\
  \end{array}\right)_{[1]} \otimes\cdots\otimes\left(\begin{array}{c}
    1 \\
    0 \\
    0 \\
  \end{array}\right)_{[N]} .\label{left-vacuum}
\eea
The operators (\ref{AB-operator}) acting on the quasi-vacuum states give rise to
\bea
&&\langle 0|\,A_i(u)=\a_i(u)\,\langle 0|,\quad i=1,2,3, \label{A-action}\\[4pt]
&&\langle 0|\,B_i(u)=0,\quad \langle 0|\,C_i(u)\neq 0,\quad i=1,2,3,\label{B-action}\\[4pt]
&&A_i(u)\,|0\rangle=\a_i(u)\,|0\rangle,\quad i=1,2,3, \label{A-action-1}\\[4pt]
&&C_i(u)\,|0\rangle=0,\quad B_i(u)\,|0\rangle\neq 0,\quad i=1,2,3,\label{B-action-1}
\eea
where the functions $\a_i(u)$ are
\bea
&&\a_1(u)=\prod_{l=1}^N c(u-\theta_l)=\prod_{l=1}^N\{\sinh(u-\theta_l-5\eta)+\sinh\eta\},\no\\
&&\a_2(u)=\prod_{l=1}^N b(u-\theta_l)=\prod_{l=1}^N\{\sinh(u-\theta_l-3\eta)+\sinh3\eta\},\no\\
&&\a_3(u)=\prod_{l=1}^N d(u-\theta_l)=\prod_{l=1}^N\{\sinh(u-\theta_l-\eta)+\sinh\eta\}.\label{a-functions}
\eea
For convenience, we introduce  two  functions
\bea
&&\hspace{-1.2truecm}
\xi(u)=e^{\eta}\frac{\a_3(u^{(2)})}{\a_2(u^{(1)})},\quad
\bar\xi(u)=e^{-\eta}\frac{\a_3(u^{(2)})}{\a_2(u^{(1)})},\label{function-xi}
\eea
where we have used the convention: $ u^{(1)}=u+4\eta\ ;u^{(2)}=u+6\eta+i\pi$.
\subsection{A convenient basis for the IK model}
In this subsection, we construct a convenient basis for the IK model, and parameterize it as follows. For  two non-negative integers $m_2$ and $m$ such that $m_2\leq m\leq N$,
let us introduce  a $m$-tuple positive integers $P=\{p_1,\cdots,p_m\}$, which  satisfy the relation
\bea
1\leq p_1<p_2<\cdots<p_{m_2}\leq N,\quad 1\leq p_{m_2+1}<\cdots<p_m\leq N,\quad  {\rm and}\quad p_j\neq p_l.\label{P-condition}
\eea
For each $P$, let us introduce  a left state $\langle \theta^{(2)}_{p_m},\cdots,\theta^{(2)}_{p_{m_2+1}};\theta^{(1)}_{p_{m_2}},\cdots,\theta^{(1)}_{p_1}|$ and a right state
$|\theta^{(1)}_{p_1},\cdots,\theta^{(1)}_{p_{m_2}};\theta^{(2)}_{p_{m_2+1}},\cdots,\theta^{(2)}_{p_m}\rangle$ parameterized by the $N$ inhomogeneity parameters $\{\theta_j\}$ as follows:
\bea
&&\hspace{-1.2truecm}\langle \theta^{(2)}_{p_m},\cdots,\theta^{(2)}_{p_{m_2+1}};\theta^{(1)}_{p_{m_2}},\cdots,\theta^{(1)}_{p_1}| =
\langle 0|C_2(\theta^{(2)}_{p_{m}})\cdots C_2(\theta^{(2)}_{p_{m_2+1}})\,C_1(\theta^{(1)}_{p_{m_2}})\cdots
C_1(\theta^{(1)}_{p_{1}}), \label{left-state}\\[4pt]
&&\hspace{-1.2truecm}|\theta^{(1)}_{p_1},\cdots,\theta^{(1)}_{p_{m_2}};\theta^{(2)}_{p_{m_2+1}},\cdots,\theta^{(2)}_{p_m}\rangle=
B_1(\theta^{(1)}_{p_{1}})\cdots B_1(\theta^{(1)}_{p_{m_2}}) \,B_2(\theta^{(2)}_{p_{m_2+1}})\cdots B_2(\theta^{(2)}_{p_{m}})|0\rangle,
\label{right-state}
\eea
where $m_2$ (resp. $m-m_2$) is the number of the operators $C_1(u)$ or $B_1(u)$
(resp. $C_2(u)$ or $B_2(u)$), and $ \theta^{(1)}_{i}=\theta_{i}+4\eta\ ;\theta^{(2)}_{i}=\theta_{i}+6\eta+i\pi$.
It is  easy to check that the states
(\ref{left-state}) and (\ref{right-state}) are common eigenstates of the operator $A_1(u)$ with different $u$, namely,
\bea
&&\hspace{-1.2truecm}\langle \theta^{(2)}_{p_m},\cdots,\theta^{(2)}_{p_{m_2+1}};\theta^{(1)}_{p_{m_2}},\cdots,\theta^{(1)}_{p_1}|
\,A_1(u)= \a_1(u) \prod_{i=1}^{m_2}z(\theta^{(1)}_{p_{i}}-u)\prod_{l=m_2+1}^{m}\frac{c(\theta^{(2)}_{p_{l}}-u)}
{d(\theta^{(2)}_{p_{l}}-u)}\no\\[2pt]
&&\quad\quad\times \langle \theta^{(2)}_{p_m},\cdots,\theta^{(2)}_{p_{m_2+1}};\theta^{(1)}_{p_{m_2}},\cdots,\theta^{(1)}_{p_1}| , \label{Eigenvalue-left-3}\\[4pt]
&&\hspace{-1.2truecm}A_1(u)\,|\theta^{(1)}_{p_1},\cdots,\theta^{(1)}_{p_{m_2}};\theta^{(2)}_{p_{m_2+1}},\cdots,\theta^{(2)}_{p_m}\rangle
= \a_1(u) \prod_{i=1}^{m_2}z(\theta^{(1)}_{p_{i}}-u)\prod_{l=m_2+1}^{m}\frac{c(\theta^{(2)}_{p_{l}}-u)}
{d(\theta^{(2)}_{p_{l}}-u)}\no\\[2pt]
&&\quad\quad\times |\theta^{(1)}_{p_1},\cdots,\theta^{(1)}_{p_{m_2}};\theta^{(2)}_{p_{m_2+1}},\cdots,\theta^{(2)}_{p_m}\rangle,\label{Eigenvalue-right-3}
\eea
where the functions $z(u)$ and $\a_i(u)$ are given by (\ref{Functions}) and (\ref{a-functions}).
From the exchange relations given by (\ref{Exchang-1})-(\ref{Exchang-2}) below, we can  verify the above relations. It is easy to show that the states
(\ref{left-state}) and (\ref{right-state}) are non-zeros thanks to the orthogonal relations (see below (\ref{Orthnormal}) and (\ref{Normal-factor})).
\subsection{Orthogonality and other properties of the basis}

With help of the exchange relations given by (\ref{Exchang-1})-(\ref{Exchang-2}) below, we can derive some quasi-symmetry properties of the left states
$\{\langle \theta^{(2)}_{p_m},\cdots,\theta^{(2)}_{p_{m_2+1}};\theta^{(1)}_{p_{m_2}},\cdots,\theta^{(1)}_{p_1}|\}$
\footnote{Similar results can also be obtained for the right states.}
\bea
&&\hspace{-1.3truecm}\langle \theta^{(2)}_{p_m},\cdots,\theta^{(2)}_{p_{m_2+1}};\theta^{(1)}_{p_{m_2}},\cdots,
\theta^{(1)}_{p_{i+1}},\theta^{(1)}_{p_{i}},\theta^{(1)}_{p_{i-1}},\cdots,\theta^{(1)}_{p_1}|
\no\\&&\hspace{-0.3truecm}
=w(\theta^{(1)}_{p_{i+1}}-\theta^{(1)}_{p_i})
\langle \theta^{(2)}_{p_m},\cdots,\theta^{(2)}_{p_{m_2+1}};\theta^{(1)}_{p_{m_2}},\cdots,
\theta^{(1)}_{p_{i+2}},\theta^{(1)}_{p_{i}},\theta^{(1)}_{p_{i+1}},\theta^{(1)}_{p_{i-1}},\cdots,\theta^{(1)}_{p_1}|,
\eea\label{quasi-symmetry1}
\bea
&&\hspace{-2.7truecm}\langle \theta^{(2)}_{p_m},\cdots,\theta^{(2)}_{p_{m_2+1}};\theta^{(1)}_{p_{m_2}},\cdots,\theta^{(1)}_{p_1}|
\no\\&&\hspace{-1.5truecm}
=z(\theta^{(2)}_{p_{m_2+1}}-\theta^{(1)}_{p_{m_2}})
\langle\theta^{(2)}_{p_m},\cdots,\theta^{(2)}_{p_{m_2+2}},\theta^{(1)}_{p_{m_2}},\theta^{(2)}_{p_{m_2+1}},
\theta^{(1)}_{p_{m_2-1}},\cdots,\theta^{(1)}_{p_1}|,
\eea\label{quasi-symmetry2}
\bea
&&\hspace{-3.4truecm}\langle \theta^{(2)}_{p_m},\cdots,\theta^{(2)}_{p_{i+1}},\theta^{(2)}_{p_{i}},\theta^{(2)}_{p_{i-1}},\cdots,
\theta^{(2)}_{p_{m_2+1}};\theta^{(1)}_{p_{m_2}},\cdots,\theta^{(1)}_{p_1}|
\no\\ &&\hspace{-2.2truecm}
=\langle \theta^{(2)}_{p_m},\cdots,
\theta^{(2)}_{p_{i+2}},\theta^{(2)}_{p_{i}},\theta^{(2)}_{p_{i+1}},\theta^{(2)}_{p_{i-1}},\cdots,
\theta^{(2)}_{p_{m_2+1}};\theta^{(1)}_{p_{m_2}},\cdots,\theta^{(1)}_{p_1}|.\label{quasi-symmetry3}
\eea
Noting the fact that $\a_1(\theta^{(i)}_l)=0$, for $l=1,\cdots,N;\ i=1,2$, we can also obtain some useful identities
\bea
&&\hspace{-1.2truecm}\langle \theta^{(2)}_{p_m},\cdots,\theta^{(2)}_{p_{m_2+1}};\theta^{(1)}_{p_{m_2}},\cdots,\theta^{(1)}_{p_1}|\,T^i_j(\theta^{(1)}_{p_l})=0,\quad T^i_j=B_1,B_2,A_1,\label{Vanshing-1}\\[4pt]
&&\hspace{-1.2truecm}\langle \theta^{(2)}_{p_m},\cdots,\theta^{(2)}_{p_{m_2+1}};\theta^{(1)}_{p_{m_2}},\cdots,\theta^{(1)}_{p_1}|\,T^i_j(\theta^{(2)}_{p_l})=0,\quad T^i_j=B_1,B_2,B_3,C_1,A_1,A_2,\label{Vanshing-2}\\[4pt]
&&\hspace{-1.2truecm}T^i_j(\theta^{(1)}_{p_l})|\theta^{(1)}_{p_1},\cdots,\theta^{(1)}_{p_{m_2}};\theta^{(2)}_{p_{m_2+1}},\cdots,\theta^{(2)}_{p_m}\rangle=0,\quad
T^i_j=C_1,C_2,A_1,\label{Vanshing-3}\\[4pt]
&&\hspace{-1.2truecm}T^i_j(\theta^{(2)}_{p_l})|\theta^{(1)}_{p_1},\cdots,\theta^{(1)}_{p_{m_2}};\theta^{(2)}_{p_{m_2+1}},\cdots,\theta^{(2)}_{p_m}\rangle=0,\quad
T^i_j=C_1,C_2,C_3,B_1,A_1,A_2.\label{Vanshing-4}
\eea
It should be emphasized that in the above identities $l$ takes the value of  $m+1,\cdots,N$.
As an example, a brief proof for the identity (\ref{Vanshing-4})  is given in Appendix B.
These properties and the exchange relations of the operators
allow us to derive the orthogonal relations between the left states
and the right states \bea
&&\hspace{-1truecm}
\langle \theta^{(2)}_{q_{m'}},\cdots,\theta^{(2)}_{q_{m'_2+1}};\theta^{(1)}_{q_{m'_2}},\cdots,\theta^{(1)}_{q_1}
|\theta^{(1)}_{p_1},\cdots,\theta^{(1)}_{p_{m_2}};\theta^{(2)}_{p_{m_2+1}},\cdots,\theta^{(2)}_{p_m}\rangle \no\\&&\quad\quad
=\delta_{m,m'}\,\delta_{m_2,m'_2}\prod_{k=1}^{m}\delta_{p_k,q_k}\,
G_m(\theta^{(1)}_{p_1},\cdots,\theta^{(1)}_{p_{m_2}};\theta^{(2)}_{p_{m_2+1}},\cdots,\theta^{(2)}_{p_m}),
\label{Orthnormal}
\eea
where the factor $G_m(\theta^{(1)}_{p_1},\cdots,\theta^{(1)}_{p_{m_2}};\theta^{(2)}_{p_{m_2+1}},\cdots,\theta^{(2)}_{p_m})$ is given by
\bea
&&\hspace{-0.8truecm}
G_m(\theta^{(1)}_{p_1},\cdots,\theta^{(1)}_{p_{m_2}};\theta^{(2)}_{p_{m_2+1}},\cdots,\theta^{(2)}_{p_m})=
\prod_{k=1}^{m_2}\left \{\right. 2\cosh\eta \sinh(2\eta)\, \a^{(1)}_{p_k}(\theta^{(1)}_{p_k})\alpha_2(\theta^{(1)}_{p_k})\no\\[4pt]
&&\times \prod_{\substack{i=1\\i\neq k}}^{m_2}z(\theta^{(1)}_{p_k}-\theta^{(1)}_{p_i}) \prod_{l=k+1}^{m_2}\omega(\theta^{(1)}_{p_l}-\theta^{(1)}_{p_k})
\prod_{j=m_2+1}^{m}\frac{c(\theta^{(2)}_{p_j}-\theta^{(1)}_{p_k})}{d(\theta^{(2)}_{p_j}-\theta^{(1)}_{p_k})}
z(\theta^{(2)}_{p_j}-\theta^{(1)}_{p_k})z(\theta^{(1)}_{p_k}-\theta^{(2)}_{p_j})
\left.\right \} \no\\[4pt]
&&\times \!\!
\prod_{k=m_2+1}^{m}\!\left\{ \right.f(0)\a^{(1)}_{p_k}(\theta^{(2)}_{p_k})\alpha_3(\theta^{(2)}_{p_k})\prod_{\substack{i=m_2+1\\i\neq k}}^{m}
\frac{c(\theta^{(2)}_{p_k}-\theta^{(2)}_{p_i})}{d(\theta^{(2)}_{p_k}-\theta^{(2)}_{p_i})}\left.\right\}. \label{Normal-factor}
\eea
The functions $\{\a_i^{(1)}(u)\}$ are
\bea
\a_i^{(1)}(u)=\prod_{\substack{k=1\\k\neq i}}^N c(u-\theta_k)
=\prod_{\substack{k=1\\k\neq i}}^N\{\sinh(u-\theta_k-5\eta)+\sinh\eta\},\quad i=1,\cdots,N. \label{d-function-1}
\eea
On the other hand, we know that the total number of the linear-independent left (right) states given in (\ref{left-state}) and  (\ref{right-state}) is
\begin{eqnarray}
\sum_{m=0}^N\frac{N!}{(N-m)!m!}\sum_{m_2=0}^m\frac{m!}{(m-m_2)!m_2!}&=&\sum_{m=0}^N\frac{N!}{(N-m)!m!}\,2^m=3^N.\nonumber
\end{eqnarray}
Thus these right (left) states
form an orthogonal right (left) basis of the Hilbert space, namely,
\bea
&&{\rm id}=\sum_{m=0}^N\sum_{m_2=0}^m\sum_{P}\frac{1}
{G_m(\theta^{(1)}_{p_1},\cdots,\theta^{(1)}_{p_{m_2}};\theta^{(2)}_{p_{m_2+1}},\cdots,\theta^{(2)}_{p_m})}\no\\[4pt]
&&\quad\quad\times |\theta^{(1)}_{p_1},\cdots,\theta^{(1)}_{p_{m_2}};\theta^{(2)}_{p_{m_2+1}},\cdots,\theta^{(2)}_{p_m}\rangle
\langle \theta^{(2)}_{p_m},\cdots,\theta^{(2)}_{p_{m_2+1}};\theta^{(1)}_{p_{m_2}},\cdots,\theta^{(1)}_{p_1}|,\label{Identity}
\eea where the notation $\sum_{P}$ indicates the sum over all possible combination $P$
satisfying the condition (\ref{P-condition}).

Some remarks are in order.  The states given by  (\ref{left-state}) (resp. (\ref{right-state}))  are  eigenstates of the commutative family $A_1(u)$  and
serve as the basis of the left (right) Hilbert space for generic inhomogeneous parameters $\{\theta_j\}$.   These kind of states are relevant to
the separation of variables (SoV) \cite{Skl92} states  and the F-basis \cite{Mai96} for the quantum spin chain associated with the $A$-type algebra.
For the $su(2)$ case, the corresponding states  are the SoV states  for the XXZ spin chain, and was shown in \cite{Nic12} that it coincides with the so-called F-basis \cite{Mai96}. For the $su(n)$ case, the corresponding states are the nested generalization of the SoV states \cite{Hao16} for the trigonometric $su(n)$ spin chain and coincide with the associated F-basis \cite{Alb00, Alb00-1, Alb01, Yan04, Zha06}.

\subsection{Operators in the basis}

The exchange relations (\ref{Exchang-1})-(\ref{Exchang-2}) and the identities (\ref{Vanshing-1})-(\ref{Vanshing-4}) enable us
to calculate the actions of the operators $A_i(u)$, $B_i(u)$ and $C_i(u)$ on the basis given by (\ref{left-state}) and (\ref{right-state}).
Direct calculation shows that the resulting actions on this basis become much simpler, comparing with those on the original basis.
Here we list some of them relevant for us to obtain the explicit expressions of Bethe states
\bea\label{Expressions-3}
&&\hspace{-1.3in}
\langle\theta^{(2)}_{p_m},\cdots,\theta^{(2)}_{p_{m_2+1}};\theta^{(1)}_{p_{m_2}},\cdots,\theta^{(1)}_{p_1}|
\,A_1(u)= \a_1(u) \prod_{i=1}^{m_2}z(\theta^{(1)}_{p_{i}}-u)\prod_{l=m_2+1}^{m}\frac{c(\theta^{(2)}_{p_{l}}-u)}
{d(\theta^{(2)}_{p_{l}}-u)}\no\\[2pt]
&&\hspace{-2.7truecm}\times \langle \theta^{(2)}_{p_m},\cdots,\theta^{(2)}_{p_{m_2+1}};\theta^{(1)}_{p_{m_2}},\cdots,\theta^{(1)}_{p_1}|,
\eea
\bea
&&\hspace{-0.5in}
\langle\theta^{(2)}_{p_m},\cdots,\theta^{(2)}_{p_{m_2+1}};\theta^{(1)}_{p_{m_2}},\cdots,\theta^{(1)}_{p_1}|\,B_1(u)=
\sum^{m_2}_{i=1}\frac{\bar e(\theta^{(1)}_{p_i}-u)}{b(\theta^{(1)}_{p_i}-u)}\a_1(u)\a_2(\theta^{(1)}_{p_i})
\prod^{i-1}_{h=1}\omega(\theta^{(1)}_{p_i}-\theta^{(1)}_{p_h})
\no \\ &&
\times\prod^{m_2}_{\substack{j=1\\j\neq i}}z(\theta^{(1)}_{p_j}-u)\frac{z(\theta^{(1)}_{p_i}-\theta^{(1)}_{p_j}) }
{\omega(\theta^{(1)}_{p_i}-\theta^{(1)}_{p_j})} \prod^{m}_{k=m_2+1}\frac{c(\theta^{(2)}_{p_k}-u)}{d(\theta^{(2)}_{p_k}-u)}
z(\theta^{(1)}_{p_i}-\theta^{(2)}_{p_k}) z(\theta^{(2)}_{p_k}-\theta^{(1)}_{p_i})\no\\ &&\times\langle\theta^{(2)}_{p_m},\cdots,
\theta^{(2)}_{p_{m_2+1}};\theta^{(1)}_{p_{m_2}},\cdots,\hat\theta^{(1)}_{p_i}
\cdots,\theta^{(1)}_{p_1}|
\no\\\no\\&&
\hspace{-0.2in}+\sum^{m}_{i=m_2+1}\bigg[  \frac{\bar e(\theta^{(2)}_{p_i}-u)}{d(\theta^{(2)}_{p_i}-u)}-\sum^{m_2}_{l=1}\frac
{e(\theta^{(1)}_{p_l}-\theta^{(2)}_{p_i})\bar e(\theta^{(1)}_{p_l}-u)c(\theta^{(2)}_{p_i}-u)}
{b(\theta^{(1)}_{p_l}-\theta^{(2)}_{p_i})b(\theta^{(1)}_{p_l}-u)d(\theta^{(2)}_{p_i}-u)}
\prod^{m_2}_{\substack{j=1\\j\neq l}}z(\theta^{(1)}_{p_j}-u)z(\theta^{(1)}_{p_l}-\theta^{(1)}_{p_j}) \bigg]
\no\\&&\times
\prod^{m}_{\substack{k=m_2+1\\k\neq i}} \frac{b(\theta^{(2)}_{p_k}-\theta^{(1)}_{p_i})c(\theta^{(2)}_{p_k}-u)}
{c(\theta^{(2)}_{p_k}-\theta^{(1)}_{p_i})d(\theta^{(2)}_{p_k}-u)}
z(\theta^{(2)}_{p_i}-\theta^{(2)}_{p_k}) \a_1(u)\bar \xi(\theta_{p_i})
\no\\&&
\times\langle\theta^{(2)}_{p_m},\cdots,\hat\theta^{(2)}_{p_i},\cdots,\theta^{(2)}_{p_{m_2+1}};
\theta^{(1)}_{p_i},\theta^{(1)}_{p_{m_2}},\cdots,\theta^{(1)}_{p_1}|,\label{B1-decomposition}
\eea
\bea
&&\hspace{-0.5in}
\langle\theta^{(2)}_{p_m},\cdots,\theta^{(2)}_{p_{m_2+1}};\theta^{(1)}_{p_{m_2}},\cdots,\theta^{(1)}_{p_1}|\,B_2(u)=
\no\\&&
\hspace{-0.15in}\sum^{m}_{l=m_2+1}\frac{\bar f(\theta^{(2)}_{p_l}-u)}{d(\theta^{(2)}_{p_l}-u)}
\prod^{m_2}_{i=1}\frac{d(\theta^{(1)}_{p_i}-u)}{b(\theta^{(1)}_{p_i}-u)}
\prod^{m}_{\substack{j=m_2+1\\j\neq l}}\frac{c(\theta^{(2)}_{p_j}-u)c(\theta^{(2)}_{p_l}-\theta^{(2)}_{p_j})}
{d(\theta^{(2)}_{p_j}-u)d(\theta^{(2)}_{p_l}-\theta^{(2)}_{p_j})}
\times\a_1(u)\a_3(\theta^{(2)}_{p_l})
\no\\&&
\times\langle\theta^{(2)}_{p_m},\cdots,\hat\theta^{(2)}_{p_l},\cdots,\theta^{(2)}_{p_{m_2+1}};\theta^{(1)}_{p_{m_2}},
\cdots,\theta^{(1)}_{p_1}|\,
\no\\\no\\&&
\hspace{-0.3in}+\sum^{m}_{l=m_2+1}\sum^{m}_{i>l}\bigg\{\frac{\bar g(\theta^{(2)}_{p_l}-u)\bar e(\theta^{(2)}_{p_i}-u)}{d(\theta^{(2)}_{p_l}-u)
d(\theta^{(2)}_{p_i}-u)}-\frac{\bar f(\theta^{(2)}_{p_l}-u)c(\theta^{(2)}_{p_i}-u)}{d(\theta^{(2)}_{p_l}-u)d(\theta^{(2)}_{p_i}-u)
\bar y(\theta^{(2)}_{p_l}-\theta^{(2)}_{p_i})}\bigg\}w(\theta^{(1)}_{p_l}-\theta^{(1)}_{p_i})
\no\\&&\times
\prod^{m}_{\substack{j=m_2+1\\j\neq l,i}} \frac{c(\theta^{(2)}_{p_j}-u)
z(\theta^{(2)}_{p_i}-\theta^{(2)}_{p_j})z(\theta^{(2)}_{p_l}-\theta^{(2)}_{p_j})}{d(\theta^{(2)}_{p_j}-u)
z(\theta^{(2)}_{p_j}-\theta^{(1)}_{p_i})z(\theta^{(2)}_{p_j}-\theta^{(1)}_{p_l})}
\prod^{m_2}_{k=1}\frac{d(\theta^{(2)}_{p_k}-u)}{b(\theta^{(2)}_{p_k}-u)}
\times\a_1(u)\bar \xi(\theta_{p_i})\bar \xi(\theta_{p_l})
\no\\&&\times
\langle\theta^{(2)}_{p_m},\cdots,\hat\theta^{(2)}_{p_i}\cdots,\hat\theta^{(2)}_{p_l},
\cdots,\theta^{(2)}_{p_{m_2+1}};\theta^{(1)}_{p_i},\theta^{(1)}_{p_l},\theta^{(1)}_{p_{m_2}},\cdots,\theta^{(1)}_{p_1}|\,
\no\\\no\\&&
\hspace{-0.3in}+\sum^{m_2}_{l=1}\sum^{m_2}_{i>l}\bigg\{\frac{\bar g(\theta^{(1)}_{p_l}-u)\bar e(\theta^{(1)}_{p_i}-u)}
{b(\theta^{(1)}_{p_l}-u)b(\theta^{(1)}_{p_i}-u)}-\frac{\bar f(\theta^{(1)}_{p_l}-u)g(\theta^{(1)}_{p_l}-\theta^{(1)}_{p_i})}
{b(\theta^{(1)}_{p_l}-u)d(\theta^{(1)}_{p_l}-\theta^{(1)}_{p_i})}z(\theta^{(1)}_{p_i}-u)\bigg\}
\a_1(u)\a_2(\theta^{(1)}_{p_l})\a_2(\theta^{(1)}_{p_i})
\no\\&&\times
\prod^{l-1}_{h=1} w(\theta^{(1)}_{p_l}-\theta^{(1)}_{p_h})
\prod^{i-1}_{\substack{h=1\\h\neq l}} w(\theta^{(1)}_{p_i}-\theta^{(1)}_{p_h})
\prod^{m_2}_{\substack{j=1\\j\neq l,i}}z(\theta^{(1)}_{p_j}-u)\frac{z(\theta^{(1)}_{p_i}-\theta^{(1)}_{p_j}) z(\theta^{(1)}_{p_l}-\theta^{(1)}_{p_j})}{w(\theta^{(1)}_{p_i}-\theta^{(1)}_{p_j})w(\theta^{(1)}_{p_l}-\theta^{(1)}_{p_j})}
\no\\&&\times
\prod^{m}_{k=m_2+1}\frac{c(\theta^{(2)}_{p_k}-u)}{d(\theta^{(2)}_{p_k}-u)}z(\theta^{(1)}_{p_l}-\theta^{(2)}_{p_k})
z(\theta^{(2)}_{p_k}-\theta^{(1)}_{p_l})z(\theta^{(1)}_{p_i}-\theta^{(2)}_{p_k}) z(\theta^{(2)}_{p_k}-\theta^{(1)}_{p_i})
\no\\&&\times
\langle\theta^{(2)}_{p_m},\cdots,\theta^{(2)}_{p_{m_2+1}};\theta^{(1)}_{p_{m_2}},\cdots,
\hat\theta^{(1)}_{p_i},\cdots,\hat\theta^{(1)}_{p_l},\cdots,\theta^{(1)}_{p_1}|\,
\no\\&&
\hspace{-0.3in}+\sum^{m_2}_{l=1}\sum^{m}_{i=m_2+1}
\Bigg\{\frac{\bar e(\theta^{(2)}_{p_i}-u)}{d(\theta^{(2)}_{p_i}-u)}
\frac{\bar g(\theta^{(1)}_{p_l}-u)z(\theta^{(1)}_{p_l}-\theta^{(1)}_{p_i})}
{b(\theta^{(1)}_{p_l}-u)w(\theta^{(1)}_{p_l}-\theta^{(1)}_{p_i})}
\prod^{m_2}_{\substack{h=1\\h\neq l}}\frac{z(\theta^{(1)}_{p_l}-\theta^{(1)}_{p_h})}
{w(\theta^{(1)}_{p_l}-\theta^{(1)}_{p_h})}
\no\\&&+
\bigg[\frac{\bar g(\theta^{(2)}_{p_i}\!-\theta^{(1)}_{p_l})}{b(\theta^{(2)}_{p_i}\!-\theta^{(1)}_{p_l})}\!-
\frac{\bar f(\theta^{(2)}_{p_i}\!-\theta^{(1)}_{p_l})}
{b(\theta^{(2)}_{p_i}\!-\theta^{(1)}_{p_l})\bar y(\theta^{(2)}_{p_i}\!-\theta^{(1)}_{p_l})}\bigg ]
\frac{\bar f(\theta^{(1)}_{p_l}\!-u)c(\theta^{(2)}_{p_i}\!-u)}{b(\theta^{(1)}_{p_l}\!-u)d(\theta^{(2)}_{p_i}\!-u)}
\prod^{m_2}_{\substack{j=1\\j\neq l}}\frac{c(\theta^{(1)}_{p_l}\!-\theta^{(1)}_{p_j})z(\theta^{(1)}_{p_j}\!-u)}
{d(\theta^{(1)}_{p_l}\!-\theta^{(1)}_{p_j})w(\theta^{(1)}_{p_l}\!-\theta^{(1)}_{p_j})}
\no\\&&+
\sum^{m_2}_{\substack{k=1\\k\neq l}}
\bigg[
\frac{g(\theta^{(1)}_{p_l}-\theta^{(1)}_{p_k})z(\theta^{(1)}_{p_l}-\theta^{(1)}_{p_k})\bar f(\theta^{(1)}_{p_l}-u)
z(\theta^{(1)}_{p_k}-u)}
{d(\theta^{(1)}_{p_l}-\theta^{(1)}_{p_k})w(\theta^{(1)}_{p_l}-\theta^{(1)}_{p_k})b(\theta^{(1)}_{p_l}-u)}
-\frac{\bar e(\theta^{(1)}_{p_k}-u)\bar g(\theta^{(1)}_{p_l}-u)}{b(\theta^{(1)}_{p_k}-u)b(\theta^{(1)}_{p_l}-u)}
\bigg]
\no\\&&\times
\frac {e(\theta^{(1)}_{p_k}-\theta^{(2)}_{p_i})c(\theta^{(2)}_{p_i}-u)z(\theta^{(1)}_{p_l}-\theta^{(1)}_{p_i})}
{b(\theta^{(1)}_{p_k}-\theta^{(2)}_{p_i})d(\theta^{(2)}_{p_i}-u)w(\theta^{(1)}_{p_l}-\theta^{(1)}_{p_i})}
\prod^{m_2}_{\substack{j=1\\j\neq l,k}}
\frac{z(\theta^{(1)}_{p_j}-u)z(\theta^{(1)}_{p_k}-\theta^{(1)}_{p_j})z(\theta^{(1)}_{p_l}-\theta^{(1)}_{p_j})}
{w(\theta^{(1)}_{p_l}-\theta^{(1)}_{p_j})}\Bigg\}
\no\\&&\times
\prod^{m}_{\substack{j=m_2+1\\j\neq i}}\frac{c(\theta^{(2)}_{p_j}-u)z(\theta^{(2)}_{p_i}-\theta^{(2)}_{p_j})}{d(\theta^{(2)}_{p_j}-u)z(\theta^{(2)}_{p_j}-\theta^{(1)}_{p_i})}
z(\theta^{(1)}_{p_l}-\theta^{(2)}_{p_j})z(\theta^{(2)}_{p_j}-\theta^{(1)}_{p_l})
\times\a_1(u)\a_2(\theta^{(1)}_{p_l})\bar \xi(\theta_{p_i})
\no\\&&
\times\langle\theta^{(2)}_{p_m},\cdots,
\hat\theta^{(2)}_{p_i},\cdots,\theta^{(2)}_{p_{m_2+1}};\theta^{(1)}_{p_i},\theta^{(1)}_{p_{m_2}},
\cdots,\hat\theta^{(1)}_{p_l},\cdots,\theta^{(1)}_{p_1}|,\label{B2-decomposition}
\eea
where the parameter with a hat {$\hat \theta^{(i)}_{p_j}$}  means this parameter is absent and the functions $\xi(u)$ and $\bar{\xi}(u)$ are given by (\ref{function-xi}).

Expanding the operators $A_i(u)$, $B_i(u)$ and $C_i(u)$ in terms of the local operators $\{|i\rangle_l\langle j|\,|\,i,j=1,2,3;\,l=1,\cdots, N\}$
(i.e., in original basis) gives rise to that the total number of all summing terms in the decomposition for  each operator may increase
exponentially with $N$ (which was shown even for the very simple case of the XXZ chain \cite{Mai96}). In contrast, the total number of
summing terms for  each decomposition in  (\ref{Expressions-3})-(\ref{B2-decomposition}) only increases  as a polynomial of $N$. This leads to the fact
that the actions of the  monodromy matrices in the very basis provided by (\ref{left-state})-(\ref{right-state})  can be simplified dramatically.
It is believed that such a basis would play the same role for the IK model as that of the F-basis for the quantum spin chains
related to the $A$-type  (super)algebras \cite{Mai96, Alb00, Alb00-1, Alb01, Yan04, Zha06}. Moreover, such  simplified actions of the
creation operators further allow us to construct the  recursive relations of the Bethe states, which uniquely determine the state.

\section{Bethe states in the basis}
\setcounter{equation}{0}

\subsection{Bethe states}
The off-shell Bethe states  of the IK model can be constructed  by the recursive relation \cite{Tar88}
\bea \label{IK bethe state}
&&\hspace{-1.5truecm}|\phi_n(u_1,\cdots,u_n)\rangle=B_1(u_1)|\phi_{n-1}(u_2,\cdots,u_n)\rangle \no\\
&&\hspace{-0.5truecm}-B_2(u_1)\sum_{i=2}^n\frac{\a_1(u_i)}{y(u_1-u_i)}\prod_{j=2}^{i-1} \omega(u_i-u_j)\prod_{\substack{k=2\\k\neq i}}^n z(u_k-u_i)|\phi_{n-2}(u_2,\cdots,\hat u_i\cdots,u_n)\rangle,
\eea
where the parameter with a hat {$\hat u_i$} means this parameter is absent and the initial conditions of the above recursive relations are
\begin{equation}
 |\phi_0\rangle=|0\rangle, \quad \quad |\phi_1(u)\rangle= B_1(u)|0\rangle.
\end{equation}
These states  become the eigenstates of the transfer matrix $t(u)$ (or on-shell) if the parameters satisfy the Bethe Ansatz equations (BAEs) \cite{Tar88}
\bea
\frac{\a_1(u_j)}{\a_2(u_j)}=
\prod^{n}_{\substack{k=1\\k\neq j}}\frac{z(u_j-u_k)}{z(u_k-u_j)}w(u_k-u_j),
\quad\quad j=1,\cdots,n.
\eea
Using (\ref{B1-decomposition}) and (\ref{B2-decomposition}) we can calculate the expressions of the Bethe states in terms of the  basis (\ref{left-state}) as follows.
Let us define  scalar products of the Bethe state  $|\phi_n(u_1,\cdots,u_n)\rangle$ with vectors in the basis
\bea\label{F-define}
&&\langle\theta^{(2)}_{p_m},\cdots,\theta^{(2)}_{p_{m_2+1}};\theta^{(1)}_{p_{m_2}},\cdots,\theta^{(1)}_{p_1}|
\phi_n(u_1,\cdots,u_n)\rangle
\no\\&&\quad\quad=
F_{2m-m_2,n}(\theta^{(2)}_{p_m},\cdots,\theta^{(2)}_{p_{m_2+1}};\theta^{(1)}_{p_{m_2}},\cdots,\theta^{(1)}_{p_1}|u_1,\cdots,u_n).\label{scalar-product}
\eea
It is easy to verify that
\bea\label{F-define-1}
&&F_{2m-m_2,n}(\theta^{(2)}_{p_m},\cdots,\theta^{(2)}_{p_{m_2+1}};\theta^{(1)}_{p_{m_2}},\cdots,\theta^{(1)}_{p_1}|u_1,\cdots,u_n)
\no\\&&\quad\quad=
\delta_{2m-m_2,n}
F_n(\theta^{(2)}_{p_m},\cdots,\theta^{(2)}_{p_{m_2+1}};\theta^{(1)}_{p_{m_2}},\cdots,\theta^{(1)}_{p_1}|u_1,\cdots,u_n).
\eea
With the help of the relations (\ref{IK bethe state}), (\ref{B1-decomposition}) and (\ref{B2-decomposition}) and following the method in \cite{Yan04}, we can derive some  recursive relations among these scalar products
\bea
&&\hspace{-1truecm}F_n(\theta^{(2)}_{p_m},\cdots,\theta^{(2)}_{p_{m_2+1}};\theta^{(1)}_{p_{m_2}},\cdots,\theta^{(1)}_{p_1}|u_1,\cdots,u_n)=
\no\\&&\hspace{0.3truecm}C_1F_{n-1}(\theta^{(2)}_{p_m},\cdots,\theta^{(2)}_{p_{m_2+1}};\theta^{(1)}_{p_{m_2}},\cdots,\hat\theta^{(1)}_{p_i}
\cdots,\theta^{(1)}_{p_1}|u_2,\cdots,u_n)
\no\\&&+C_2F_{n-1}(\theta^{(2)}_{p_m},\cdots,\hat\theta^{(2)}_{p_i},\cdots,\theta^{(2)}_{p_{m_2+1}};
\theta^{(1)}_{p_i},\theta^{(1)}_{p_{m_2}},\cdots,\theta^{(1)}_{p_1}|u_2,\cdots,u_n)
\no\\&&+C_3F_{n-2}(\theta^{(2)}_{p_m},\cdots,\hat\theta^{(2)}_{p_l},\cdots,\theta^{(2)}_{p_{m_2+1}};\theta^{(1)}_{p_{m_2}},
\cdots,\theta^{(1)}_{p_1}|u_2,\cdots,\hat u_j\cdots,u_n)
\no\\&&+C_4F_{n-2}(\theta^{(2)}_{p_m},\cdots,\hat\theta^{(2)}_{p_i}\cdots,\hat\theta^{(2)}_{p_l},
\cdots,\theta^{(2)}_{p_{m_2+1}};\theta^{(1)}_{p_i},\theta^{(1)}_{p_l},\theta^{(1)}_{p_{m_2}},\cdots,\theta^{(1)}_{p_1}
|u_2,\cdots,\hat u_j\cdots,u_n)
\no\\&&+C_5F_{n-2}(\theta^{(2)}_{p_m},\cdots,\theta^{(2)}_{p_{m_2+1}};\theta^{(1)}_{p_{m_2}},\cdots,
\hat\theta^{(1)}_{p_i},\cdots,\hat\theta^{(1)}_{p_l},\cdots,\theta^{(1)}_{p_1}|u_2,\cdots,\hat u_j\cdots,u_n)
\no\\&&+C_6F_{n-2}(\theta^{(2)}_{p_m},\cdots,
\hat\theta^{(2)}_{p_i},\cdots,\theta^{(2)}_{p_{m_2+1}};\theta^{(1)}_{p_i},\theta^{(1)}_{p_{m_2}},
\cdots,\hat\theta^{(1)}_{p_l},\cdots,\theta^{(1)}_{p_1}|u_2,\cdots,\hat u_j\cdots,u_n),\label{Recursive-relation}
\eea
where the concrete form of the coefficients $C_i(i=1,\cdots,6)$ are given in Appendix C. The above recursive relation allows one to determine
each scalar products in (\ref{scalar-product})  uniquely.
Here we give the explicit expressions of the  first two $F_1$ and $F_2$ of the functions
\bea
&&\hspace{-1.1truecm}F_1(\theta^{(1)}_{p_1}|u_1)=\frac{\bar e(\theta^{(1)}_{p_1}-u_1)}{b(\theta^{(1)}_{p_1}-u_1)}
\a_1(u_1)\a_2(\theta^{(1)}_{p_1}),
\no\\\no\\&&
\hspace{-1.1truecm}F_2(\theta^{(2)}_{p_1}|u_1,u_2)=
\bigg[\frac{\bar e(\theta^{(2)}_{p_1}-u_1)\bar e(\theta^{(1)}_{p_1}-u_2)}
{d(\theta^{(2)}_{p_1}-u_1)b(\theta^{(1)}_{p_1}-u_2)}e^{-\eta}
-\frac{\bar f(\theta^{(2)}_{p_1}-u_1)}{y(u_1-u_2)d(\theta^{(2)}_{p_1}-u_1)} \bigg]
\a_1(u_1)\a_1(u_2)\a_3(\theta^{(2)}_1),
\no\\[4pt]\no\\&&
\hspace{-1.1truecm}F_2(\theta^{(1)}_{p_2},\theta^{(1)}_{p_1}|u_1,u_2)=\bigg\{
\frac{\bar e(\theta^{(1)}_{p_1}-u_1)z(\theta^{(1)}_{p_1}-\theta^{(1)}_{p_2})\bar e(\theta^{(1)}_{p_2}-u_2)
z(\theta^{(1)}_{p_2}-u_1)}
{b(\theta^{(1)}_{p_1}-u_1)w(\theta^{(1)}_{p_1}-\theta^{(1)}_{p_2})b(\theta^{(1)}_{p_2}-u_2)}
\no\\[4pt]&&
+
\frac{\bar e(\theta^{(1)}_{p_2}-u_1)z(\theta^{(1)}_{p_2}-\theta^{(1)}_{p_1})\bar e(\theta^{(1)}_{p_1}-u_2)z(\theta^{(1)}_{p_1}-u_1)}
{b(\theta^{(1)}_{p_2}-u_1)b(\theta^{(1)}_{p_1}-u_2)}
+
\frac{z(\theta^{(1)}_{p_2}-u_1)\bar f(\theta^{(1)}_{p_1}-u_1)g(\theta^{(1)}_{p_1}-\theta^{(1)}_{p_2})}
{y(u_1-u_2)b(\theta^{(1)}_{p_1}-u_1)d(\theta^{(1)}_{p_1}-\theta^{(1)}_{p_2})}
\no\\[4pt]&&
-
\frac{\bar g(\theta^{(1)}_{p_1}-u_1)\bar e(\theta^{(1)}_{p_1}-u_2)}
{y(u_1-u_2)b(\theta^{(1)}_{p_1}-u_1)b(\theta^{(1)}_{p_2}-u_1)}\bigg\}
\a_1(u_1)\a_1(u_2)\a_2(\theta^{(1)}_{p_1})\a_2(\theta^{(1)}_{p_2}).
\eea
According to (\ref{Identity}), (\ref{F-define}) and (\ref{F-define-1}), we can expand the Bethe states (\ref{IK bethe state}) as
\bea
&&\hspace{-1.5truecm}|\phi_n(u_1,\cdots,u_n)\rangle={\rm id}\times |\phi_n(u_1,\cdots,u_n)\rangle
\no \\ [4pt] && \hspace{-0.8truecm}
=\sum_{m=0}^N\sum_{m_2=0}^m\sum_{P}\frac{1}
{G_m(\theta^{(1)}_{p_1},\cdots,\theta^{(1)}_{p_{m_2}};\theta^{(2)}_{p_{m_2+1}},\cdots,\theta^{(2)}_{p_m})}
|\theta^{(1)}_{p_1},\cdots,\theta^{(1)}_{p_{m_2}};\theta^{(2)}_{p_{m_2+1}},\cdots,\theta^{(2)}_{p_m}\rangle
\no\\[4pt]&&\quad \times
\langle \theta^{(2)}_{p_m},\cdots,\theta^{(2)}_{p_{m_2+1}};\theta^{(1)}_{p_{m_2}},\cdots,\theta^{(1)}_{p_1}
|\phi_n(u_1,\cdots,u_n)\rangle
\no \\[4pt]&&\hspace{-0.8truecm}
=\sum_{2m-m_2=n}\sum_{P}
\frac{F_n(\theta^{(2)}_{p_m},\cdots,\theta^{(2)}_{p_{m_2+1}};\theta^{(1)}_{p_{m_2}},\cdots,\theta^{(1)}_{p_1}|u_1,\cdots,u_n)}
{G_m(\theta^{(1)}_{p_1},\cdots,\theta^{(1)}_{p_{m_2}};\theta^{(2)}_{p_{m_2+1}},\cdots,\theta^{(2)}_{p_m})}
\no\\[4pt]&&\quad \times
|\theta^{(1)}_{p_1},\cdots,\theta^{(1)}_{p_{m_2}};\theta^{(2)}_{p_{m_2+1}},\cdots,\theta^{(2)}_{p_m}\rangle \label{Bethe-state-E},
\eea
 where the notation $\sum_{2m-m_2=n}$ indicates the sum over all integers $\{0\leq m_2\leq m\leq N\}$
satisfying the condition: $2m-m_2=n$. Thanks to the fact that the scalar products $F_i(u)$ defined by (\ref{scalar-product}) can be
determined by the very recursive relations (\ref{Recursive-relation}). This allows us to give the explicit expressions of the Bethe
states of the IK model with the periodic boundary condition.

\subsection{Inverse Problem}
The important problem in the theory of quantum integrable models, after diagonalizing the corresponding Hamiltonians, is to solve
the corresponding quantum inverse scattering problem. Namely, local  operators are reconstructed in terms of the matrix elements
of the monodromy-matrix. The general method to solve the problem for a quantum integrable spin chain was given in \cite{Goh00,Mai00}.
It is easy to check that the $R$-matrix (\ref{R-matrix})-(\ref{R-element-2}) of the IK model possesses the required  properties:
\bea
&&\hspace{-1.5cm}\mbox{ Initial
condition}:\,R_{12}(0)= \varphi P_{12},\label{Int-R}\\
&&\hspace{-1.5cm}\mbox{ Unitarity
relation}:\,R_{12}(u)R_{21}(-u)= \phi(u)\,\times {\rm id},\label{Unitarity}\no
\eea
where
\bea
\varphi=\sinh\eta-\sinh5\eta,\quad \phi(u)=[\sinh\eta+\sinh(u-5\eta)][\sinh\eta-\sinh(u+5\eta)]. \no
\eea
The $P_{ij}$ is the permutation operator.
As shown in \cite{Mai00}, these properties of the $R$-matrix directly indicate the identity:
\bea
tr_0\lt\{x_{0}\,T_0(\theta_i)\rt\}=\prod^{i-1}_{j=1}t^{-1}(\theta_{j})\,x_{i}\,\prod^{i}_{j=1}t(\theta_{j}),\no
\eea
where $\{x_j|j=0,\cdots,N\}$ are local operators acting on the j-th space and  $t(u)$ is the transfer matrix. Define the local operator $e^{ij}=|i\rangle \langle j|, 1\leq i,j\leq3$
and let $x_0=e^{ij}_0$, and then we can express the local spin operators in terms of the operator entries of the monodromy-matrix. As an example, here we list some of them
\begin{eqnarray}\label{inverse}
   e^{11}_i\!\!&=&\prod^{N-1}_{i=1}\prod^{N}_{j>i}\phi^{-1}_{ij}\varphi^{-N} \prod^{i-1}_{j=1}t(\theta_{j})\,A_1(\theta_i)\,\prod^{N}_{j=i+1}t(\theta_{j}), \no \\
   e^{12}_i\!\!&=&\prod^{N-1}_{i=1}\prod^{N}_{j>i}\phi^{-1}_{ij}\varphi^{-N} \prod^{i-1}_{j=1}t(\theta_{j})\,C_1(\theta_i)\,\prod^{N}_{j=i+1}t(\theta_{j}), \no \\   e^{13}_i\!\!&=&\prod^{N-1}_{i=1}\prod^{N}_{j>i}\phi^{-1}_{ij}\varphi^{-N} \prod^{i-1}_{j=1}t(\theta_{j})\,C_2(\theta_i)\,\prod^{N}_{j=i+1}t(\theta_{j}), \no \\   e^{21}_i\!\!&=&\prod^{N-1}_{i=1}\prod^{N}_{j>i}\phi^{-1}_{ij}\varphi^{-N} \prod^{i-1}_{j=1}t(\theta_{j})\,B_1(\theta_i)\,\prod^{N}_{j=i+1}t(\theta_{j}), \no \\   e^{22}_i\!\!&=&\prod^{N-1}_{i=1}\prod^{N}_{j>i}\phi^{-1}_{ij}\varphi^{-N} \prod^{i-1}_{j=1}t(\theta_{j})\,A_2(\theta_i)\,\prod^{N}_{j=i+1}t(\theta_{j}), \no \\   e^{31}_i\!\!&=&\prod^{N-1}_{i=1}\prod^{N}_{j>i}\phi^{-1}_{ij}\varphi^{-N} \prod^{i-1}_{j=1}t(\theta_{j})\,B_2(\theta_i)\,\prod^{N}_{j=i+1}t(\theta_{j}), \no \\     e^{33}_i\!\!&=&\prod^{N-1}_{i=1}\prod^{N}_{j>i}\phi^{-1}_{ij}\varphi^{-N} \prod^{i-1}_{j=1}t(\theta_{j})\,A_3(\theta_i)\,\prod^{N}_{j=i+1}t(\theta_{j}),
\end{eqnarray}
where
\bea
\phi_{ij}\!\!=[\sinh\eta+\sinh(\theta_i-\theta_j-5\eta)][\sinh\eta-\sinh(\theta_i-\theta_j+5\eta)].\no
\eea

Due to the fact that the Bethe states (\ref{IK bethe state})  are obtained by acting the creation operators $B_1(u)$ and $B_2(u)$ (for the left Bethe state, by the acting the creation operators $C_1(u)$ and $C_2(u)$) on
the corresponding reference state and that  all the local operators $\{e^{ij}_l|l=1,\cdots,N\}$  have been reconstructed in terms of the operators  $A_i(u)$, $B_i(u)$ and $C_i(u)$ as (\ref{inverse}), one can perform the corresponding F-basis analysis of correlation functions \cite{Kor93} of the IK model like those in the quantum integrable spin chains associated with the $A$-type (super)algebras \cite{Kit99, Ter99, Zha06}.


\section{Conclusions}
We have introduced  a convenient  basis (\ref{left-state}) and (\ref{right-state}) of the Hilbert space for the IK model with the periodic boundary condition, which is the quantum spin chain
associated with the $A^{(2)}_{2}$ algebra. It is shown that matrix elements of the monodromy matrix  acting
on the very  basis take simple forms (\ref{Expressions-3})-(\ref{B2-decomposition}), which is quite similar as that in
the F-basis for a quantum spin chain associated with $A$-type (super)algebra.
As an application, we have obtained the  recursive relations (\ref{Recursive-relation})  of vector components of the Bethe states of the model  in the very  basis,
which allow us uniquely to determine the states. With the explicit expressions (\ref{Bethe-state-E}) of the Bethe states and the solution of quantum inverse problem, one can further
calculate the correlation functions of the IK model with the periodic boundary condition.

It is well-known \cite{Res83} that taking the rational limit (i.e.,$\eta\rightarrow0$) the IK model becomes the $su(3)$-invariant spin chain. It is easy to show that in this limit the resulting basis of (\ref{left-state}) and (\ref{right-state}) is exactly the rational version of the basis given recently in \cite{Hao16} which coincides with  the F-basis \cite{Alb00}  of the $su(3)$-invariant closed chain.

\section*{Acknowledgments}

We would like to thank Prof. Y. Wang for his valuable discussions and continuous encouragements. The financial supports from the National Program for Basic Research of MOST (Grant Nos. 2016YFA0300600 and 2016YFA0302104), the National Natural Science Foundation of China (Grant Nos. 11434013, 11425522, 11547045, 11774397, 11775178 and 11775177), the Major Basic Research Program of Natural Science of Shaanxi Province (Grant Nos. 2017KCT-12, 2017ZDJC-32) and the Strategic Priority Research Program of the Chinese Academy of Sciences are gratefully acknowledged. Y. Qiao is also supported by the NWU graduate student innovation funds No. YYB17003.

\section*{Appendix A: Exchange relations}
\renewcommand{\baselinestretch}{1.6}
The quadratic commutation relation (\ref{RLL})   allows us to derive the exchange relations among the operators (\ref{T-matrix})  given by the matrix elements
of the momodromy matrix $T(u)$.  Here we list some of the exchange relations among the monodromy-matrix elements
which have been  used in our calculation
\setcounter{equation}{0}
\renewcommand{\theequation}{A.\arabic{equation}}
\begin{footnotesize}
\begin{spacing}{0.5}
\begin{eqnarray}
  [A_1(u),\,A_1(v)]=[B_2(u),\,B_2(v)]=[C_2(u),\,C_2(v)]=[A_3(u),\,A_3(v)]=0
,\label{Exchang-1}\end{eqnarray}
\begin{eqnarray}\label{ER-B1-B1}
  B_1(u)B_1(v)=w(v-u)\bigg[B_1(v)B_1(u)-\frac{1}{y(v-u)}B_2(v)A_1(u)\bigg]+\frac{1}{y(u-v)}
  B_2(u)A_1(v)
,\end{eqnarray}
\begin{eqnarray}
  A_1(u)B_1(v)=z(v-u)B_1(v)A_1(u)-\frac{e(v-u)}{b(v-u)}B_1(u)A_1(v)
,\end{eqnarray}
\begin{eqnarray}
  A_1(u)B_2(v)=\frac{c(v-u)}{d(v-u)}B_2(v)A_1(u)-\frac{g(v-u)}{d(v-u)}
  B_1(u)B_1(v)-\frac{f(v-u)}{d(v-u)}B_2(u)A_1(v)
,\end{eqnarray}
\begin{eqnarray}\label{ER-B1-B2}
  B_1(u)B_2(v)=z(v-u)B_2(v)B_1(u)-\frac{e(v-u)}{b(v-u)}B_2(u)B_1(v)
,\end{eqnarray}
\begin{eqnarray}
  B_2(u)B_1(v)=\frac{1}{z(u-v)}B_1(v)B_2(u)+\frac{e(u-v)}{c(u-v)}B_2(v)B_1(u)
,\end{eqnarray}
\begin{eqnarray}
  C_1(u)B_1(v)=B_1(v)C_1(u)-\frac{e(v-u)}{b(v-u)}\bigg[A_2(u)A_1(v)-A_2(v)A_1(u)\bigg]
,\end{eqnarray}
\begin{eqnarray}
  B_1(u)B_3(v)=B_3(v)B_1(u)+\frac{\bar e(v-u)}{b(v-u)}B_2(v)A_2(u)-\frac{e(v-u)}{b(v-u)}B_2(u)A_2(v)
,\end{eqnarray}
\begin{eqnarray}
  B_2(u)B_3(v)=\frac{1}{z(v-u)}B_3(v)B_2(u)+\frac{\bar e(v-u)}{c(v-u)}B_2(v)B_3(u)
,\end{eqnarray}
\begin{eqnarray}
  B_3(u)B_2(v)=\frac{1}{z(v-u)}B_2(v)B_3(u)+\frac{e(v-u)}{b(v-u)}B_3(v)B_2(u)
,\end{eqnarray}
\begin{equation}
\hspace{-0.2in}  A_2(u)B_2(v)\!=\!z(u-v)z(v-u)B_2(v)A_2(u)+\frac{\bar e(u-v)}{b(u-v)}\bigg[
  B_3(u)B_1(v)-B_1(u)B_3(v)+\frac{\bar e(u-v)}{b(u-v)}B_2(u)A_2(v)\bigg]
,\end{equation}
\begin{eqnarray}
\hspace{-0.4in}  A_3(u)B_1(v)\!=\!\frac{b(u-v)}{d(u-v)}B_1(v)A_3(u)-\frac{1}{y(u-v)}
  B_3(u)A_2(v)+\frac{e(u-v)}{d(u-v)}B_2(v)C_3(u)-\frac{\bar f(u-v)}{d(u-v)}B_2(u)C_3(v)
,\end{eqnarray}
\begin{eqnarray}
  A_3(u)B_2(v)\!=\!\frac{c(u-v)}{d(u-v)}B_2(v)A_3(u)-\frac{1}{y(u-v)}B_3(u)B_3(v)
  -\frac{\bar f(u-v)}{d(u-v)}B_2(u)A_3(v)
,\end{eqnarray}
\begin{eqnarray}
\hspace{-0.4in}  C_1(u)B_2(v)\!=\!\frac{b(v-u)}{d(v-u)}B_2(v)C_1(u)+\frac{e(v-u)}{d(v-u)}
  B_3(v)A_1(u)-\frac{f(v-u)}{d(v-u)}B_3(u)A_1(v)-\frac{g(v-u)}{d(v-u)}A_2(u)B_1(v)
,\end{eqnarray}
\begin{eqnarray}
\hspace{-0.4in}  C_3(u)B_2(v)\!=\!\frac{d(v-u)}{b(v-u)}B_2(v)C_3(u)+\frac{g(v-u)}{b(v-u)}
  B_3(v)A_2(u)+\frac{f(v-u)}{b(v-u)}A_3(v)B_1(u)-\frac{e(v-u)}{b(v-u)}A_3(u)B_1(v)
,\end{eqnarray}
\begin{eqnarray}
\hspace{-0.4in}  C_2(u)B_1(v)\!=\!\frac{d(v-u)}{b(v-u)}B_1(v)C_2(u)+\frac{g(v-u)}{b(v-u)}
  A_2(v)C_1(u)+\frac{f(v-u)}{b(v-u)}C_3(v)A_1(u)-\frac{e(v-u)}{b(v-u)}C_3(u)A_1(v)
,\end{eqnarray}
\begin{eqnarray}
\hspace{-0.4in}  C_3(u)A_1(v)\!=\!\frac{b(u-v)}{d(u-v)}A_1(v)C_3(u)+\frac{e(u-v)}{d(u-v)}
  B_1(v)C_2(u)-\frac{1}{y(u-v)}A_2(u)C_1(v)-\frac{\bar f(u-v)}{d(u-v)}B_1(u)C_2(v)
,\end{eqnarray}
\begin{eqnarray}
\hspace{-0.4in}  C_2(u)B_2(v)\!\!=\!\!B_2(v)C_2(u)+\frac{1}{\bar y(v-u)}\bigg[
  B_3(v)C_1(u)-C_3(u)B_1(v)\bigg]\!+\!\frac{f(v-u)}{d(v-u)}\bigg[
  A_3(v)A_1(u)-A_3(u)A_1(v)\bigg]
,\end{eqnarray}
\begin{eqnarray}
\hspace{-0.4in}C_2(u)B_2(v)\!\!=\!\!B_2(v)C_2(u)+\frac{1}{y(u-v)}\bigg[
  B_1(v)C_3(u)-C_1(u)B_3(v)\bigg]\!+\!\frac{\bar f(u-v)}{d(u-v)}\bigg[
  A_1(v)A_3(u)-A_1(u)A_3(v)\bigg]
,\end{eqnarray}
\begin{eqnarray}
  C_1(u)B_3(v)\!\!&=&\frac{a(v-u)}{d(v-u)}B_3(v)C_1(u)+\frac{g(v-u)}{d(v-u)}
  A_3(v)A_1(u)+\frac{1}{y(v-u)}B_2(v)C_2(u)\no \\&
  -&\frac{g(v-u)}{d(v-u)}A_2(u)A_2(v)-\frac{f(v-u)}{d(v-u)}B_3(u)C_1(v)
,\end{eqnarray}
\begin{eqnarray}
  A_2(u)B_1(v)\!\!&=&\frac{z(u-v)}{w(u-v)}B_1(v)A_2(u)-\frac{\bar e(u-v)}{b(u-v)}
  B_1(u)A_2(v)+\frac{1}{y(u-v)}B_3(u)A_1(v)\no \\&
  +&\frac{\bar e(u-v)}{y(u-v)b(u-v)}B_2(u)C_1(v)-\frac{z(u-v)}{w(u-v)y(u-v)}B_2(v)C_1(u)
,\end{eqnarray}
\begin{eqnarray}
  C_3(u)B_1(v)\!\!&=&\frac{a(u-v)}{d(u-v)}B_1(v)C_3(u)+\frac{g(u-v)}{d(u-v)}
  B_2(v)C_2(u)-\frac{\bar f(u-v)}{d(u-v)}B_1(u)C_3(v)\no \\&
  +&\frac{1}{y(u-v)}\bigg[A_1(v)A_3(u)-A_2(u)A_2(v)\bigg]
.\label{Exchang-2}\end{eqnarray}
\end{spacing}
\end{footnotesize}

\section*{Appendix B: Proof of the vanishing properties}
\renewcommand{\baselinestretch}{1.6}
\setcounter{equation}{0}
\renewcommand{\theequation}{B.\arabic{equation}}
As a typical example, we give a brief proof of the identity (\ref{Vanshing-4}), namely,
\begin{equation}\label{B1-bra}
B_1(\theta^{(2)}_{p_l})|\theta^{(1)}_{p_1},\cdots,\theta^{(1)}_{p_{m_2}};\theta^{(2)}_{p_{m_2+1}},\cdots,\theta^{(2)}_{p_m}\rangle=0,
\quad l=m+1,\cdots,N.
\end{equation}
We prove the above identity  by the induction. First we need to prove
\begin{equation}\label{B1-bra-0}
  B_1(\theta^{(2)}_{p_l})|0\rangle=0.
\end{equation}
It is easy to check that it is true for the $N=1$ case. The proof goes by induction in the number of particles starting from $N=1$.
Assume that (\ref{B1-bra-0}) were also true for the cases of $N=1,\cdots,L$, which can be denoted as
\begin{equation}
  B_1^{L}(\theta^{(2)}_{p_l})|0\rangle^{L}=0.\label{B-formal}
\end{equation}
where the operator $X^{L}$ means $X$ is embedded in the $L$ tensor space.
We show that it is valid for $N=L+1$
\begin{eqnarray}
   B_1^{L+1}(\theta^{(2)}_{p_l})|0\rangle^{L+1}\!&\!=\!&\!\bigg\{
   A_1^{(L+1)}(\theta^{(2)}_{p_l}) \otimes B_1^{L}(\theta^{(2)}_{p_l})
  +B_1^{(L+1)}(\theta^{(2)}_{p_l}) \otimes A_2^{L}(\theta^{(2)}_{p_l})\no \\&&
  +B_2^{(L+1)}(\theta^{(2)}_{p_l}) \otimes C_3^{L}(\theta^{(2)}_{p_l})\bigg\}
  \bigg\{|0\rangle^{(L+1)} \otimes |0\rangle^{L}\bigg\}\no \\\!&\!
  =\!&\!0,
\end{eqnarray}
where the operator $X^{(L+1)}$ means $X$ is embedded in the $(L+1)$-th space.
Thus, the relation  (\ref{B1-bra-0}) is proven.
Using (\ref{B1-bra-0}) and the exchange relations (\ref{ER-B1-B2}), we can easily get
\begin{equation}\label{B1-bra-1}
B_1(\theta^{(2)}_{p_l})|\theta^{(2)}_{p_{m_2+1}},\cdots,\theta^{(2)}_{p_m}\rangle=0.
\end{equation}
Finally, the exchange relations (\ref{ER-B1-B1}) and (\ref{B1-bra-1}) allow us to derive
\begin{eqnarray}
&&\hspace{0.3truecm}B_1(\theta^{(2)}_{p_l})|\theta^{(1)}_{p_1},\cdots,\theta^{(1)}_{p_{m_2}};
\theta^{(2)}_{p_{m_2+1}},\cdots,\theta^{(2)}_{p_m}\rangle
\no \\&&\quad\quad=
\prod^{m_2}_{j=1}w(\theta^{(1)}_{p_j}-\theta^{(2)}_{p_l})
B_1(\theta^{(1)}_{p_1}),\cdots,B_1(\theta^{(1)}_{p_{m_2}})B_1(\theta^{(2)}_{p_l})
|\theta^{(2)}_{p_{m_2+1}},\cdots,\theta^{(2)}_{p_m}\rangle
\no \\&&\quad\quad=0.
\end{eqnarray}
Thus, the relation  (\ref{B1-bra}) has been proved.

\section*{Appendix C: Coefficients of the recursive relation (\ref{Recursive-relation})}
\setcounter{equation}{0}
\renewcommand{\theequation}{C.\arabic{equation}}

By using the relations (\ref{B1-decomposition}), (\ref{B2-decomposition}) and (\ref{IK bethe state}), we can derive the recursive relations (\ref{Recursive-relation})
among the functions $\{F_i(u)\}$, with their coefficients being given as follows:
\bea
&&\hspace{-3.3truecm}
C_1\!=
\sum^{m_2}_{i=1}\frac{\bar e(\theta^{(1)}_{p_i}-u_1)}{b(\theta^{(1)}_{p_i}-u_1)}
\prod^{i-1}_{h=1}\omega(\theta^{(1)}_{p_i}-\theta^{(1)}_{p_h})
\prod^{m_2}_{\substack{j=1\\j\neq i}}z(\theta^{(1)}_{p_j}-u_1)\frac{z(\theta^{(1)}_{p_i}-\theta^{(1)}_{p_j}) }
{\omega(\theta^{(1)}_{p_i}-\theta^{(1)}_{p_j})}\times\a_1(u_1)\a_2(\theta^{(1)}_{p_i})
\no\\&&\hspace{-2.3truecm}\times\!\!
\prod^{m}_{k=m_2+1}\frac{c(\theta^{(2)}_{p_k}-u_1)}{d(\theta^{(2)}_{p_k}-u_1)}
z(\theta^{(1)}_{p_i}-\theta^{(2)}_{p_k}) z(\theta^{(2)}_{p_k}-\theta^{(1)}_{p_i}),
\eea
\bea
&&\hspace{-1.2truecm}
C_2\!=\!\!
\sum^{m}_{i=m_2+1}\bigg[\frac{\bar e(\theta^{(2)}_{p_i}\!-u_1)}{d(\theta^{(2)}_{p_i}\!-u_1)}-\sum^{m_2}_{l=1}\frac
{e(\theta^{(1)}_{p_l}-\theta^{(2)}_{p_i})\bar e(\theta^{(1)}_{p_l}-u_1)c(\theta^{(2)}_{p_i}-u_1)}
{b(\theta^{(1)}_{p_l}-\theta^{(2)}_{p_i})b(\theta^{(1)}_{p_l}-u_1)d(\theta^{(2)}_{p_i}-u_1)}
\prod^{m_2}_{\substack{j=1\\j\neq l}}z(\theta^{(1)}_{p_j}\!-u_1)z(\theta^{(1)}_{p_l}\!-\theta^{(1)}_{p_j}) \bigg]
\no\\&&\hspace{-0.2truecm}\times\!\!
\prod^{m}_{\substack{k=m_2+1\\k\neq i}} \frac{b(\theta^{(2)}_{p_k}-\theta^{(1)}_{p_i})c(\theta^{(2)}_{p_k}-u_1)}
{c(\theta^{(2)}_{p_k}-\theta^{(1)}_{p_i})d(\theta^{(2)}_{p_k}-u_1)}
z(\theta^{(2)}_{p_i}-\theta^{(2)}_{p_k})\times\a_1(u_1)\bar \xi(\theta_{p_i}),
\eea
\bea
&&\hspace{-2truecm}
C_3\!=\!
-\!\sum^{m}_{l=m_2+1}\frac{\bar f(\theta^{(2)}_{p_l}-u_1)}{d(\theta^{(2)}_{p_l}-u_1)}\prod^{m}_
{\substack{j=m_2+1\\j\neq l}}\frac{c(\theta^{(2)}_{p_j}-u_1)c(\theta^{(2)}_{p_l}-\theta^{(2)}_{p_j})}
{d(\theta^{(2)}_{p_j}-u_1)d(\theta^{(2)}_{p_l}-\theta^{(2)}_{p_j})}
\prod^{m_2}_{i=1}\frac{d(\theta^{(1)}_{p_i}-u_1)}{b(\theta^{(1)}_{p_i}-u_1)}\times\a_1(u_1)\a_3(\theta^{(2)}_{p_l})
\no\\&&\hspace{-1.1truecm}\times
\bigg\{\sum_{j=2}^n\frac{\a_1(u_j)}{y(u_1-u_j)}\prod_{i=2}^{j-1} \omega(u_j-u_i)
\prod_{\substack{k=2\\k\neq j}}^n z(u_k-u_j)\bigg\},
\eea
\bea
&&\hspace{-1.2truecm}
C_4\!=\!
-\!\!\sum^{m}_{l=m_2+1}\sum^{m}_{i>l}\bigg\{\frac{\bar g(\theta^{(2)}_{p_l}-u_1)\bar e(\theta^{(2)}_{p_i}-u_1)}{d(\theta^{(2)}_{p_l}-u_1)
d(\theta^{(2)}_{p_i}-u_1)}-\frac{\bar f(\theta^{(2)}_{p_l}-u_1)c(\theta^{(2)}_{p_i}-u_1)}{d(\theta^{(2)}_{p_l}-u_1)d(\theta^{(2)}_{p_i}-u_1)
\bar y(\theta^{(2)}_{p_l}-\theta^{(2)}_{p_i})}\bigg\}w(\theta^{(1)}_{p_l}-\theta^{(1)}_{p_i})
\no\\&&\hspace{-0.3truecm}\times
\prod^{m}_{\substack{j=m_2+1\\j\neq l,i}} \frac{c(\theta^{(2)}_{p_j}-u_1)
z(\theta^{(2)}_{p_i}-\theta^{(2)}_{p_j})z(\theta^{(2)}_{p_l}-\theta^{(2)}_{p_j})}{d(\theta^{(2)}_{p_j}-u_1)
z(\theta^{(2)}_{p_j}-\theta^{(1)}_{p_i})z(\theta^{(2)}_{p_j}-\theta^{(1)}_{p_l})}
\prod^{m_2}_{k=1}\frac{d(\theta^{(2)}_{p_k}-u_1)}{b(\theta^{(2)}_{p_k}-u_1)}
\times\a_1(u_1)\bar \xi(\theta_{p_i})\bar \xi(\theta_{p_l})
\no\\&&\hspace{-0.3truecm}\times
\bigg\{\sum_{j=2}^n\frac{\a_1(u_j)}{y(u_1-u_j)}\prod_{i=2}^{j-1} \omega(u_j-u_i)
\prod_{\substack{k=2\\k\neq j}}^n z(u_k-u_j)\bigg\},
\eea
\bea
&&\hspace{-1.2truecm}
C_5\!=\!
-\sum^{m_2}_{l=1}\sum^{m_2}_{i>l}\bigg\{\frac{\bar g(\theta^{(1)}_{p_l}-u_1)\bar e(\theta^{(1)}_{p_i}-u_1)}
{b(\theta^{(1)}_{p_l}-u_1)b(\theta^{(1)}_{p_i}-u_1)}-\frac{\bar f(\theta^{(1)}_{p_l}-u_1)g(\theta^{(1)}_{p_l}-\theta^{(1)}_{p_i})}
{b(\theta^{(1)}_{p_l}-u_1)d(\theta^{(1)}_{p_l}-\theta^{(1)}_{p_i})}z(\theta^{(1)}_{p_i}-u_1)\bigg\}
\no\\&&\hspace{-0.3truecm}\times
\prod^{l-1}_{h=1} w(\theta^{(1)}_{p_l}-\theta^{(1)}_{p_h})
\prod^{i-1}_{\substack{h=1\\h\neq l}} w(\theta^{(1)}_{p_i}-\theta^{(1)}_{p_h})
\prod^{m_2}_{\substack{j=1\\j\neq l,i}}z(\theta^{(1)}_{p_j}-u_1)\frac{z(\theta^{(1)}_{p_i}\!-\theta^{(1)}_{p_j})
 z(\theta^{(1)}_{p_l}\!-\theta^{(1)}_{p_j})}{w(\theta^{(1)}_{p_i}\!-\theta^{(1)}_{p_j})w(\theta^{(1)}_{p_l}\!-\theta^{(1)}_{p_j})}
\no\\&&\hspace{-0.3truecm}\times
\prod^{m}_{k=m_2+1}\frac{c(\theta^{(2)}_{p_k}-u_1)}{d(\theta^{(2)}_{p_k}-u_1)}z(\theta^{(1)}_{p_l}-\theta^{(2)}_{p_k})
z(\theta^{(2)}_{p_k}-\theta^{(1)}_{p_l})z(\theta^{(1)}_{p_i}-\theta^{(2)}_{p_k}) z(\theta^{(2)}_{p_k}-\theta^{(1)}_{p_i})
\!\times\!\a_1(u_1)\a_2(\theta^{(1)}_{p_l})\a_2(\theta^{(1)}_{p_i})
\no\\&&\hspace{-0.3truecm}\times
\bigg\{\sum_{j=2}^n\frac{\a_1(u_j)}{y(u_1-u_j)}\prod_{i=2}^{j-1} \omega(u_j-u_i)
\prod_{\substack{k=2\\k\neq j}}^n z(u_k-u_j)\bigg\},
\eea
\bea
&&\hspace{-1.2truecm}
C_6\!=\!
-\sum^{m_2}_{l=1}\sum^{m}_{i=m_2+1}
\Bigg\{\frac{\bar e(\theta^{(2)}_{p_i}-u_1)}{d(\theta^{(2)}_{p_i}-u_1)}
\frac{\bar g(\theta^{(1)}_{p_l}-u_1)z(\theta^{(1)}_{p_l}-\theta^{(1)}_{p_i})}
{b(\theta^{(1)}_{p_l}-u_1)w(\theta^{(1)}_{p_l}-\theta^{(1)}_{p_i})}
\prod^{m_2}_{\substack{h=1\\h\neq l}}\frac{z(\theta^{(1)}_{p_l}-\theta^{(1)}_{p_h})}
{w(\theta^{(1)}_{p_l}-\theta^{(1)}_{p_h})}
\no\\&&\hspace{-0.3truecm} +
\bigg[\frac{\bar g(\theta^{(2)}_{p_i}\!-\theta^{(1)}_{p_l})}{b(\theta^{(2)}_{p_i}\!-\theta^{(1)}_{p_l})}\!-
\frac{\bar f(\theta^{(2)}_{p_i}\!-\theta^{(1)}_{p_l})}
{b(\theta^{(2)}_{p_i}\!-\theta^{(1)}_{p_l})\bar y(\theta^{(2)}_{p_i}\!-\theta^{(1)}_{p_l})}\bigg ]
\frac{\bar f(\theta^{(1)}_{p_l}\!-u_1)c(\theta^{(2)}_{p_i}\!-u_1)}{b(\theta^{(1)}_{p_l}\!-u_1)d(\theta^{(2)}_{p_i}\!-u_1)}
\prod^{m_2}_{\substack{j=1\\j\neq l}}\frac{c(\theta^{(1)}_{p_l}\!-\theta^{(1)}_{p_j})z(\theta^{(1)}_{p_j}\!-u_1)}
{d(\theta^{(1)}_{p_l}\!-\theta^{(1)}_{p_j})w(\theta^{(1)}_{p_l}\!-\theta^{(1)}_{p_j})}
\no\\&&\hspace{-0.3truecm}+
\sum^{m_2}_{\substack{k=1\\k\neq l}}
\bigg[
\frac{g(\theta^{(1)}_{p_l}-\theta^{(1)}_{p_k})z(\theta^{(1)}_{p_l}-\theta^{(1)}_{p_k})\bar f(\theta^{(1)}_{p_l}-u_1)
z(\theta^{(1)}_{p_k}-u_1)}
{d(\theta^{(1)}_{p_l}-\theta^{(1)}_{p_k})w(\theta^{(1)}_{p_l}-\theta^{(1)}_{p_k})b(\theta^{(1)}_{p_l}-u_1)}
-\frac{\bar e(\theta^{(1)}_{p_k}-u_1)\bar g(\theta^{(1)}_{p_l}-u_1)}{b(\theta^{(1)}_{p_k}-u_1)b(\theta^{(1)}_{p_l}-u_1)}
\bigg]
\no\\&&\hspace{-0.3truecm}\times
\frac {e(\theta^{(1)}_{p_k}-\theta^{(2)}_{p_i})c(\theta^{(2)}_{p_i}-u_1)z(\theta^{(1)}_{p_l}-\theta^{(1)}_{p_i})}
{b(\theta^{(1)}_{p_k}-\theta^{(2)}_{p_i})d(\theta^{(2)}_{p_i}-u_1)w(\theta^{(1)}_{p_l}-\theta^{(1)}_{p_i})}
\prod^{m_2}_{\substack{j=1\\j\neq l,k}}
\frac{z(\theta^{(1)}_{p_j}-u_1)z(\theta^{(1)}_{p_k}-\theta^{(1)}_{p_j})z(\theta^{(1)}_{p_l}-\theta^{(1)}_{p_j})}
{w(\theta^{(1)}_{p_l}-\theta^{(1)}_{p_j})}\Bigg\}
\no\\&&\hspace{-0.3truecm}\times
\prod^{m}_{\substack{j=\!m_2+1\\j\neq i}}\frac{c(\theta^{(2)}_{p_j}-u_1)z(\theta^{(2)}_{p_i}-\theta^{(2)}_{p_j})}{d(\theta^{(2)}_{p_j}-u_1)z(\theta^{(2)}_{p_j}-\theta^{(1)}_{p_i})}
z(\theta^{(1)}_{p_l}-\theta^{(2)}_{p_j})z(\theta^{(2)}_{p_j}-\theta^{(1)}_{p_l})
\times\a_1(u_1)\a_2(\theta^{(1)}_{p_l})\bar \xi(\theta_{p_i})
\no\\&&\hspace{-0.3truecm}\times
\bigg\{\sum_{j=2}^n\frac{\a_1(u_j)}{y(u_1-u_j)}\prod_{i=2}^{j-1} \omega(u_j-u_i)
\prod_{\substack{k=2\\k\neq j}}^n z(u_k-u_j)\bigg\},
\eea
where the functions $\xi(u)$ and $\bar{\xi}(u)$ are given by (\ref{function-xi}).


\end{document}